\documentclass[a4,aps]{revtex4}
\usepackage{amssymb}
\usepackage{graphicx}
\usepackage{graphics}
\usepackage{dcolumn}
\usepackage{bm}
\usepackage{color}
\usepackage{epstopdf}
\usepackage{hyperref}

\begin{document}

\newcommand{\blue}[1]{\textcolor{blue}{#1}}
\newcommand{\new}{\blue}

\title{An approach to the leading Regge pole}

\author{S. D. Campos}\email{sergiodc@ufscar.br}

\address{Universidade Federal de S\~ao Carlos, CCTS/DFQM, Sorocaba, S\~ao Paulo CEP 18052780, Brazil}
\vspace{10pt}
%\begin{indented}
%\item[]October 2019
%\end{indented}

\begin{abstract}
Neglecting spin effects, one introduces here a subtle approximation for the scattering angle, which allows the obtaining of a logarithmic leading Regge pole, consistent with the Froissart-Martin bound. A simple parameterization is also introduced for the proton-proton total cross section. Fitting procedures are implemented only for the highest energy experimental data available. The intercept for a linear approach is obtained, indicating the presence of a soft pomeron. The Tsallis entropy in the impact parameter space is calculated using the Regge pole formalism. This entropy depends on a free parameter, whose value implies the existence of a central or non-central maximum value for the entropy. The hollowness effect is discussed in terms of this parameter.
\end{abstract}

\maketitle

%%%%%%%%%%%%%%%%%%%%%%%%%%%%%%%%%%%%%%%%%%%%%%%%%%%%%%%%%%%%%%%%%%%%%%%%
\section{Introduction}\label{sec:intro}

The complex angular momentum theory or Regge theory, for short, is one of the attempts to understand the strong interaction of elementary particles initiated at the end of 1950's \cite{t.regge.nuovo.cim.14.951.1959,t.regge.nuovo.cim.18.947.1960,a.bottino.a.m.longhoni.t.regge.nuvo.cim.23.954.1962}. The straight connection between the spectrum of particles and their scattering pattern at high energy is one of the achievements of Regge theory. In the so-called Chew-Frautschi plot \cite{G.F.Chew.S.C.Frautschi.Phys.Rev.Lett.7.394.1961,G.F.Chew.S.C.Frautschi.Phys.Rev.Lett.8.41.1962}, one has an example of Regge trajectories subject to experimental verification.

The Regge poles, also called reggeons, can be divided into pomerons and odderons. The pomerons have a $C=+1$ parity and the odderons a $C=-1$ parity. Then, the pomerons have the same coupling with particles and antiparticles. The odderon distinguishes particles from antiparticles. In the scattering amplitude at very large energy ($\sqrt{s}$) and small transferred momentum ($\sqrt{|t|}$), the pomeron is the leading contribution and odderon, its counterpart, is the non-leading (or secondary) contribution \cite{S.Donnachie.book.2002}. A possible odderon exchange at small $t$ could explain the dip region in the proton-proton ($pp$) and proton-antiproton ($p\bar{p}$) scattering. In the total cross section ($\sigma_{tot}(s)$) case, the general belief is that a triple pole exchange may explain the Froissart-Martin bound saturation, $\sigma_{tot}(s)\sim(\ln s)^2$. On the other hand, the odderon contributes with a less rapid rise  ($\ln s$)  for the total cross section as $s\rightarrow\infty$. Therefore, for example, for the ISR energies, the mixing contributions of pomerons and odderons are important to take into account the experimental data subtleties. On the other hand, at LHC scale, only the pomeron contribution is dominant \cite{L.Jenkovszky.R.Schicker.I.Szanyi.Int.J.Mod.Phys.E27.1830005.2018}. Despite this theoretical understanding, several unfruitful attempts have been performed to associate a specific trajectory to the pomeron. Recently, possible experimental evidence for the odderon \cite{G.Antchev.etal.TOTEM.Coll.Eur.Phys.J.C79.785.2019} has been subject to some debate \cite{E.Martynov.B.Nicolescu.Phys.Lett.B778.414.2018,V.A.Khoze.A.D.Martin.M.G.Ryskin.Phys.Lett.B780.352.2018,A.Szczurek.P.Lebiedowicz.PoS.DIS2019.071.2019}.

The leading Regge pole, unfortunately, does not obey the Froissart-Martin bound, a remarkable theoretical result of the 1960s. Of course, one can argue this violation occurs far from the present-day energies, in the so-called Planck scale \cite{S.Donnachie.book.2002}. One can also claim this violation may be avoided through the eikonalization of the Regge pole \cite{J.L.Cardy.Nucl.Phys.B28.455-475.1971}.

In the present work, the validity of the Froissart-Martin bound in the Regge theory is restored introducing an approximation, near the forward region, resulting in logarithmic leading Regge poles. One introduces an approximation in the scattering angle representation, reducing its amplitude but allowing its logarithmic representation. Based on this result, one proposes a simple parameterization for the $pp$ total cross section. Carrying out a fitting procedure for the total cross section, considering only experimental data above 1 TeV up to the cosmic-ray data, one obtains an intercept favoring the soft pomeron. The $p\bar{p}$ experimental data are also used in the fitting procedures, as shall be explained in the text. %\new{One cannot help noticing there is a theoretical point of view where the pomeron can be considered a fractal dimension.  Then, the internal arrangement of quarks and gluons inside the hadron has fractal topology \cite{f.s.borcsik.s.d.campos_mod_phys_lett_a31_1650066_2016}.}

The geometrical arrangement of quarks and gluons, the so-called partons, brings the question of how they contribute to the hadron internal entropy and how this entropy can be utilized to explain the stability/instability of the hadrons. The formal relation between the Tsallis entropy and the real part of the profile function, in the impact parameter formalism, can help to understand how this scenario may lead to the appearance of instabilities regions inside the proton, above some critical temperature \cite{S.D.Campos.C.V.Moraes.V.A.Okorokov.Phys.Scrip.2019}. Using the results achieved here, one presents the Tsallis entropy for the logarithmic leading Regge pole in the impact parameter space. This entropy depends on a free parameter, which choice may result in a central or non-central value for the maximum of the entropy. If this parameter is independent of the impact parameter $b$, then the entropy is mostly central ($b = 0$ fm). On the other hand, if this parameter is $b$-dependent, then the entropy reaches its maximum at $b\neq 0$ fm. The $b$-dependence or $b$-independence lead to different values for the maximum of the entropy. Then, the correct choice of this parameter may help to understand the emergence of a gray area \cite{dremin_1,dremin_2}, also known as the hollowness effect \cite{broniowski_arriola,alkin_martynov,anisovich_nikonov,troshin_tyurin_1,anisovich,troshin_tyurin_2,albacete_sotoontoso,arriola_broniowski}. If the entropy is maximum at $b=0$ fm, then the hollowness effect is absent for $pp$ and $p\bar{p}$ elastic scattering. Then the maximum of the inelastic overlap function, for example, occurs at $b=0$ fm. Yet, if the maximum of the entropy occurs in $b\neq 0$ fm, then this result may favor the existence of the hollowness effect since the maximum of the inelastic overlap function also arises in $b\neq 0$ fm.

The paper is organized as follows. In section \ref{sec:rgt},  using an approximation for the scattering angle, one introduces a logarithmic leading Regge pole obeying the Froissart-Martin bound. One also presents a fitting procedure for the $pp$ total cross section above 1.0 TeV, including high energy non-accelerators data, obtaining the pomeron intercept. In a second fitting procedure, the experimental data for the $p\bar{p}$ total cross section above 1.0 TeV were also used simultaneously with the data for $pp$, resulting in a slightly different soft pomeron value. In section \ref{sec:ter}, a possible relation between the Regge pole and the Tsallis entropy is presented. The hollowness effect is also discussed based on a free parameter. Section \ref{sec:critical} presents the critical remarks.

%Recently, the proton internal distribution pressure was measured \cite{burkert_nature_vol_557_396_2018,p.e.shanahan.w.detmold.phys.rev.lett.122.072003.2019} resulting two main features: a huge value for the pressure distribution above the neutron star pressure as well as the emergence of a negative pressure distribution shielding a core with positive pressure  distribution. These experimental results were used to propose a novel picture of the proton internal arrangement, based on a geometrical view of the quark ($q$) and antiquark ($\bar{q}$) pair creation inside the proton \cite{S.D.Campos.2019}. 
%%%%%%%%%%%%%%%%%%%%%%%%%%%%%%%%%%%%%%%%%%%%%%%%%%%%%%%%%%%%%%%%%%%%%%%%

%%%%%%%%%%%%%%%%%%%%%%%%%%%%%%%%%%%%%%%%%%%%%%%%%%%%%%%%%%%%%%%%%%%%%%%%
\section{The Regge Pole and the Froissart-Martin Bound}\label{sec:rgt}

The Regge theory is an interesting way to furnish a physical interpretation of the mathematical poles that arise in the complex angular momentum plane. The scattering amplitude, in this formalism, can be written as an analytic function of the angular momentum $J$. Then, its behavior in the $s$-channel is due to the exchange of one-particle (or due to a composite particle), represented in the $t$-channel as the Regge pole. Hereafter, $s$ is the squared energy, and $t$ the squared transferred momentum, both in the center-of-mass system.

\subsection{Basic Picture}

As well-known, the Regge poles appear from the theoretical approach used and, for $s>\!\!>|t|$ ($t$ physical is negative), only the leading pole becomes important for the scattering amplitude \cite{P.D.B.Collins.Phys.Rep.4.103.1971}. Indeed, one assumes that partial wave expansion of the scattering amplitude is dominated by a finite number of isolated moving poles in the complex angular momentum plane. Mathematically, for the simple pole and a scattering angle given by 
\begin{eqnarray}\label{eq:angle}
\cos \theta=1+\frac{2t}{s},
\end{eqnarray}

\noindent one writes the following asymptotic function for the scattering amplitude
\begin{eqnarray}\label{eq:regge_pole}
A(s,t)\rightarrow (\eta+e^{-i\pi\alpha(t)})\beta(t)(s/s_0)^{\alpha(t)}, ~~~s\rightarrow\infty,
\end{eqnarray}

\noindent where $\eta=\pm 1$ is the signature related with the crossing symmetry $s\leftrightarrow u$ (or $s\leftrightarrow -s$ for high energies), $\sqrt{s_0}$ is some critical energy, and $\beta(t)$ is the residue function of the pole depending only on $t$. The asymptotic behavior of (\ref{eq:regge_pole}) comes from the fact that, for $s>\!\!>|t|$, one has the asymptotic property
\begin{eqnarray}\label{eq:gamma}
P_l(s/s_0)\rightarrow (s/s_0)^l,
\end{eqnarray}

\noindent i.e. the Legendre polynomial $P_l(s/s_0)$ is dominated by its leading term $(s/s_0)^l$. The main role in the leading Regge pole determination is performed by $\alpha(t)$, also known as the Regge trajectory. It can be obtained using the plot of mass$^2\times J$ (or spin). For example, for several mesons, it can be extracted from the experimental data collected over the years \cite{PDG-PhysRev-D98-030001-2018}. A reaction with positive $t$ implies the existence of physical particles of angular momentum $J_i$ and mass $\alpha(t_i=m_i^2)=J_i$. The $\alpha(t)$ has a simple mathematical form for positive $t$
\begin{eqnarray}\label{eq:alpha_1}
\alpha(t)=\alpha(0)+\alpha't,
\end{eqnarray}

\noindent where $\alpha'$ is the slope. The linearity of $\alpha(t)$, established a long time ago, is not true everywhere \cite{A.J.G.Hey.R.L.Kelly.Phys.Rep.96.71.1983}, and seems to be more evident for light baryons and mesons. Nonetheless, one maintains here the linearity of $\alpha(t)$ by its simplicity. Then, $\alpha(0)$ and $\alpha'$ can be easily obtained from fitting procedures of the experimental data. Using (\ref{eq:regge_pole}), one can write the differential cross section of the process as
\begin{eqnarray}\label{eq:dif_1}
\frac{d\sigma}{dt}\approx (s/s_0)^{2\alpha(t)},
\end{eqnarray}

\noindent and the resulting total cross section is given by
\begin{eqnarray}\label{eq:totalregge}
\sigma_{tot}(s)\approx (s/s_0)^{\alpha(0)-1}.
\end{eqnarray}

Despite its very successful beginning, the Regge theory does not have a precise explanation of what is the physical meaning of a Regge cut \cite{S.Donnachie.book.2002}. In general, the Regge cut is interpreted as the simultaneous exchange of two or more particles \cite{P.D.B.Collins.Phys.Rep.4.103.1971}. Moreover, the major part of its comprehension is based on the perturbation theory \cite{S.Mandelstam.Nuovo.Cim.30.1127.1963}.

At the beginning of the 1960s, the asymptotic result (\ref{eq:totalregge}) should imply that $\alpha(0)<1$, since the experimental data for the $pp$ and $p\bar{p}$ total cross section seemed to vanish or, at least, to be constant as the collision energy rise. Furthermore, the arrival of the Froissart-Martin bound in the theoretical battlefield did not contradict the decreasing observed in the experimental data for the total cross section since it runs as an upper bound and not as a mathematical limit. Thus, the result (\ref{eq:totalregge}) had sound reasonable at that time. The first version of the Pomeranchuk theorem stated that a scattering process should vanish  (asymptotically) if the charge exchange had occurred \cite{I.Ia.Pomeranchuk.Sov.Phys.JETP.7.499.1958}, in full agreement with the experimental data available. The Pomeranchuk theorem initial version states, roughly speaking, that $\sigma_{tot}^{pp}-\sigma_{tot}^{p\bar{p}}\rightarrow 0$, as the energy tends to infinity \cite{I.Ia.Pomeranchuk.Sov.Phys.JETP.7.499.1958}. Other versions, contrarily, had shown that $\sigma_{tot}^{pp}/\sigma_{tot}^{p\bar{p}}\rightarrow 1$ as $s\rightarrow\infty$ \cite{R.J.Eden.Phys.Rev.Lett.16.39.1966,G.Grunberg.T.N.Truong.Phys.Rev.Lett.31.63.1973}. In any one of these versions, one assumes (not yet been proven) that total cross section is finite for $s\rightarrow\infty$. This theoretical statement implies the existence of an exchange particle that not differs particle from antiparticle at high energies with $C=+1$. This exchange particle is the so-called pomeron, and the general belief expects to find it in the leading pole, possessing the vacuum quantum numbers and positive signature \cite{A.Donnachie.P.V.Landshoff.Phys.Lett.B.296.227.1992}. Therefore, assuming the pomeron is the leading exchange particle, then the total cross section for the $pp$ and $p\bar{p}$ scattering, for example, tends to the same value as $s\rightarrow\infty$. Table \ref{tab:table_1} shows the expected pomeron quantum numbers \cite{P.D.B.Collins.book.1977}. Without any doubt, the detection of particles with vacuum quantum numbers is a hard task in particle physics.

The odderon is defined as the $C$-odd (or $C=-1$) contribution for the scattering amplitude. At very large $s$ and small $t$, it is a non-leading contribution. It may either do not vanish relative to the pomeron or vanish as a small power of $s$ or power of $\ln s$ \cite{S.Donnachie.book.2002}. Yet, considering energies $\sqrt{s}\lesssim 1.0$ TeV, the mixing of the leading and non-leading trajectories may explain the different patterns observed in the $pp$ and $p\bar{p}$ total cross sections as well as in the dip structure for small $t$. Then, at the LHC energies, the leading contributions are dominant \cite{L.Jenkovszky.R.Schicker.I.Szanyi.Int.J.Mod.Phys.E27.1830005.2018}. Recently, it has been found a possible experimental evidence for a $C$-odd 3-gluon compound state exchange (the odderon), explained by the incompatibilities between $pp$ and $p\bar{p}$ differential cross section at $\sqrt{s}=13.0$ TeV \cite{G.Antchev.etal.TOTEM.Coll.Eur.Phys.J.C79.785.2019}. This odderon is obtained as a solution of the BFKL equation \cite{J.Bartels.L.Lipatov.G.Vacca.Phys.Lett.B477.178.2000, J.Bartels.etal.Eur.Phys.J.C.Part.Fields.20.323.2001}. In this sense, this experimental evidence corroborates the work of Jenkovszky et al. \cite{L.L.Jenkovszky.A.I.Lengyel.D.I.Lontkovskyi.Int.J.Mod.Phys.A26.4755.2011}, i.e. the observed differences between $pp$ and $p\bar{p}$ differential cross sections are only possible assuming the exchange of a $C$-odd. There is some debate about this result in the literature \cite{E.Martynov.B.Nicolescu.Phys.Lett.B778.414.2018,V.A.Khoze.A.D.Martin.M.G.Ryskin.Phys.Lett.B780.352.2018,A.Szczurek.P.Lebiedowicz.PoS.DIS2019.071.2019}.

\begin{table}[h!]
\caption{\label{tab:table_1}Expected pomeron quantum numbers which coincides with the vacuum quantum numbers. $Q$ - charge; $I$ - isospin; $S$ - strangeness; $B$ - baryon quantum number; $P$ - parity; $C$- charge conjugation; $\tau$ - signature.}
%\begin{indented}
%\item[]
\begin{tabular}{ c| c| c| c| c| c| c }
\hline
 ~Q ~& ~ I ~& ~S ~& ~B~ & P & C & $\tau$  \\ 
\hline
 0 & 0 & 0 & 0 & +1 & + 1 & + 1 \\  
\hline
\end{tabular}
%\end{indented}
\end{table}

The rise of $\sigma_{tot}^{pp}$, as $s$ increases, is an experimental fact, confirmed by the ISR in 1973. However, in the Regge theory, this growing total cross section leads to $\alpha(0)>1$, resulting in an unexpected violation of the Froissart-Martin bound. This bound was firstly obtained by Froissart \cite{m.froissart.phys.rev.123.1053.1961} in the context of Mandelstam representation, being rigorously proven through analytic functions by Martin \cite{a.martin.nuovo.cim.42.930.1966}. It can be written as \cite{a.martin.phys.rev.d80.065013.2009}
\begin{eqnarray}\label{eq:froissartmartin}
\sigma_{tot}(s)\leq \frac{1}{2m_\pi^2}\ln(s/s_0)^2.
\end{eqnarray}

Nowadays, there are several candidates to be the pomeron, $\alpha(0)=\alpha_{\mathbb{P}}(0)$. The so-called soft pomeron, constructed from multiperipheral hadronic exchanges, is the simplest one and possess an intercept $\alpha_{\mathbb{P}}^s=1.06$ \cite{P.D.B.Collins.F.D.Gault.A.Martin.Phys.Lett.B47.171.1973,R.C.Badatya.P.K.Patnaik.Pramana.15.463-474.1980} or $\alpha_{\mathbb{P}}^s=1.08$ \cite{A.Donnachie.P.V.Landshoff.Phys.Lett.B.296.227.1992}. Despite its mathematical simplicity, its dynamical origin is not well-understood \cite{H.G.Dosch.E.Ferreira.A.Kramer.Phys.Rev.D.5.1994.1992}. The hard pomeron possesses origin in the Hera and ZEUS experiments on deep inelastic scattering at DESY. The rise at $x<0.01$, where $x$ is the Bjorken scale, suggests a higher intercept $\alpha_{\mathbb{P}}^h=1.44$ \cite{A.Donnachie.P.V.Landshoff.Phys.Lett.B.518.63.2001}. However, the soft pomeron is a pole, and the hard pomeron is a cut in the $\alpha_{\mathbb{P}}(t)$ plane, i.e. an exchange of, at least, two pomerons. There are, also, the perturbative-QCD pomerons: the Low-Nussinov pomeron \cite{F.E.Low.Phys.Rev.D.12.163.1975,S.Nussinov.Phys.Rev.D.14.246.1976} and the BFKL pomeron \cite{E.A. Kuraev.L.N.Lipatov.V.S.Fadin.Sov.Phys.JETP.45.199.1977,I.I.Balitsky.L.N.Lipatov.Sov.J.Nucl.Phys.28.822.1978}. 

\subsection{Alternative Approach}

There is a mathematical disagreement between (\ref{eq:totalregge}) and (\ref{eq:froissartmartin}) for $1<\alpha_{\mathbb{P}}(0)$. The validity of the Regge theory relative to the Froissart-Martin bound can be restored imposing a constraint on the scattering angle as well as a mathematical approximation on the cosine series. Firstly, one restricts the scattering angle (\ref{eq:angle}) to the range $0\leq \cos \theta \leq 1$ as well as $|t|<\!\!< s$. This restriction covers both the forward and the near-forward scattering cases. In this range, it is possible to write the following approximation for the cosine of the scattering angle
\begin{eqnarray}\label{eq:cos_2}
\cos (\theta) = 1+\frac{2t}{s} \approx \ln \left(1+\sqrt{e}\left(1+\frac{2t}{s}\right)\right).
\end{eqnarray}
%and the Rodriguez formula given below
%\begin{eqnarray}\label{eq:rodriguez}
%P_l(x) = \frac{1}{2^l l!}\left(\frac{d}{dx}\right)^l(x^2 - 1)^l,
%\end{eqnarray}

The above result can be achieved by using the usual cosine series written as
\begin{eqnarray}\label{cos_1}
\cos(x)=1-\sum_{n=1}^{\infty} \frac{(-1)^{n-1}}{(2n)!}x^{2n}.
\end{eqnarray}

The factorial number is the problematic term for our purposes. To circumvent it, one uses Stirling's approximation
\begin{eqnarray}\label{stirling}
(2n)!\approx (2n)^{2n+\frac{1}{2}}e^{-2n}\sqrt{2\pi},
\end{eqnarray}

\noindent moreover, one notices the following inequalities are valid for $n\in \mathbb{N}$
\begin{eqnarray}\label{inel}
(2n)^{2n+\frac{1}{2}}\geq n^{n+\frac{1}{2}}\geq n^n\geq n.
\end{eqnarray}

Of course, the result (\ref{inel}), when replaced in the convenient series, possibly implies a slower convergence than the original cosine series. Now, to reduce the series on the r.h.s. of (\ref{cos_1}) into the logarithmic series, one notices that using (\ref{inel}) one can always find a real number $a$, satisfying
\begin{eqnarray}\label{inel_2}
\frac{e^{2n}}{\sqrt{2\pi}(2n)^{2n+\frac{1}{2}}}\leq \frac{a^{2n}}{n}.
\end{eqnarray}

The above inequality holds for $a=e$, for example. Using the results (\ref{stirling}) and (\ref{inel_2}), one can exchange the series on r.h.s. of (\ref{cos_1})  by the approximation
\begin{eqnarray}\label{cos_2}
\sum_{n=1}^{\infty} \frac{(-1)^{n-1}}{(2n)!}x^{2n}\approx \sum_{n=1}^{\infty}\frac{(-1)^{n-1}}{n}[(ax)^2]^n=\sum_{n=1}^{\infty}\frac{(-1)^{n-1}}{n}[(y)]^n,
\end{eqnarray}

\noindent with the condition $y\geq 0$. Using the last result, one finally obtains
\begin{eqnarray}\label{final_1}
1-\sum_{n=1}^{\infty}\frac{(-1)^{n-1}}{n}[(y)]^n=\ln\left(\frac{e}{1+y}\right).
\end{eqnarray}

To ensure the first-order approximation, one uses
\begin{eqnarray}\label{usoy}
y=e-\left[1+\sqrt{e}\left(1+\frac{2t}{s}\right)\right],
\end{eqnarray}

\noindent where the factor $\sqrt{e}$ comes from the fact that $y\geq0$. Observe that, for $t=0$, one has $\cos\theta= 1$, and the approximation performed furnish $\cos\theta\approx 0.97$. 

Using the asymptotic properties of the Legendre polynomial, one writes taking into account the approximation (\ref{eq:cos_2}),
\begin{eqnarray}\label{eq:asymp_1}
P_l(s)\rightarrow \ln (s/s_c)^{l},
\end{eqnarray}

\noindent and one adopts $\sqrt{s_c}\approx 25.0$ GeV and $\sqrt{s_c}<\sqrt{s}$ as the energy where the total cross section data achieves its minimum value \cite{S.D.Campos.C.V.Moraes.V.A.Okorokov.Phys.Scrip.2019,f.s.borcsik.s.d.campos_mod_phys_lett_a31_1650066_2016}. Indeed, the above result can be used to write the asymptotic scattering amplitude (neglecting the signature and the residue function)
\begin{eqnarray}\label{eq:asymp_2}
A(s,t)\rightarrow \ln(s/s_c)^{\alpha(t)},
\end{eqnarray}

\noindent and, therefore, the optical theorem culminate in the simple relation for the total cross section (without subtractions)
\begin{eqnarray}\label{eq:asymp_3}
\sigma_{tot}(s)\rightarrow \ln(s/s_c)^{\alpha_{\mathbb{P}}(0)}.
\end{eqnarray}

In the specified range for $\cos(\theta)$, it respects the Froissart-Martin bound if $\alpha_{\mathbb{P}}(0)\leq 2$, providing a physical relation between the pomeron intercept, $\alpha_{\mathbb{P}}(0)$, and the saturation of that bound. The particle allowing the maximum growth for $\sigma_{tot}$, obeying the Froissart-Martin bound as $s\rightarrow\infty$, is a pomeron with an intercept $\alpha_{\mathbb{P}}(0)=2$.

\subsection{Fitting Results}

As well-known, in the usual formulation of the Regge pole, the soft pomeron occurs for an intercept 1.04$\sim$1.08. Considering a total cross section written as (\ref{eq:totalregge}), this value represents the over-saturation of the Froissart-Martin bound. However, in the approach presented here, the Froissart-Martin bound saturates only for $\alpha_{\mathbb{P}}(0)\rightarrow 2$. %This constraint on the intercept may lead to a very hard pomeron, above the usual hard pomeron picture \cite{D.M.Rodrigues.E.F.Capossoli.H.Boschi-Filho.Phys.Rev.D.95.076011.2017}.

As an exercise, one extracts here $\alpha_{\mathbb{P}}(0)=\alpha_{\mathbb{P}}$ by a fitting procedure using the experimental data for $\sigma_{tot}^{pp}$ and $\sigma_{tot}^{p\bar{p}}$  above $\sqrt{s}=1.0$ TeV up to the cosmic-ray data. The experimental data were collected from Particle Data Group \cite{PDG-PhysRev-D98-030001-2018}. Moreover, $\sigma_{tot}^{pp}$ at $\sqrt{s}=2.76$ TeV is from \cite{G_Antchev_TOTEM_ Coll_Eur_Phys_J_C79_103_2019}. In this energy regime, one uses a parameterization based on the Regge approach performed above, writing
\begin{eqnarray}\label{eq:sig_tot_1}
\sigma_{tot}(s)=\beta \ln(s/s_c)^{\alpha_{\mathbb{P}}},
\end{eqnarray}

\noindent where $\beta$ and $\alpha_{\mathbb{P}}$ are free fitting parameters. This parameterization (\ref{eq:sig_tot_1}) may represents a double, $\ln(s/s_c)$, or a triple, $\ln(s/s_c)^2$, pole exchange, depending on the value of the pomeron intercept. For $\alpha_{\mathbb{P}}\rightarrow 1$, the double exchange is favored and for $\alpha_{\mathbb{P}}\rightarrow 2$, the triple pole dominates. Then, a triple pole exchange favors the saturation of the Froissart-Martin bound.

The fitting procedure is described as follows. Firstly, only the experimental data for $\sigma_{tot}^{pp}$ above 1.0 TeV are used in the fitting procedure (set 1). Secondly, the experimental data for $p\bar{p}$, above $\sqrt{s}=1.0$ TeV, are added to the $pp$ experimental data and this joint data-set is fitted (set 2). The main goal of set 2 is to compensate for the absence of $pp$ experimental data in the range $1.0$ TeV$<\sqrt{s}<2.0$ TeV, trying to improve the statistical results. One notes there is no data selection in any ensemble.

\begin{figure}
\centering{
\includegraphics[scale=0.3]{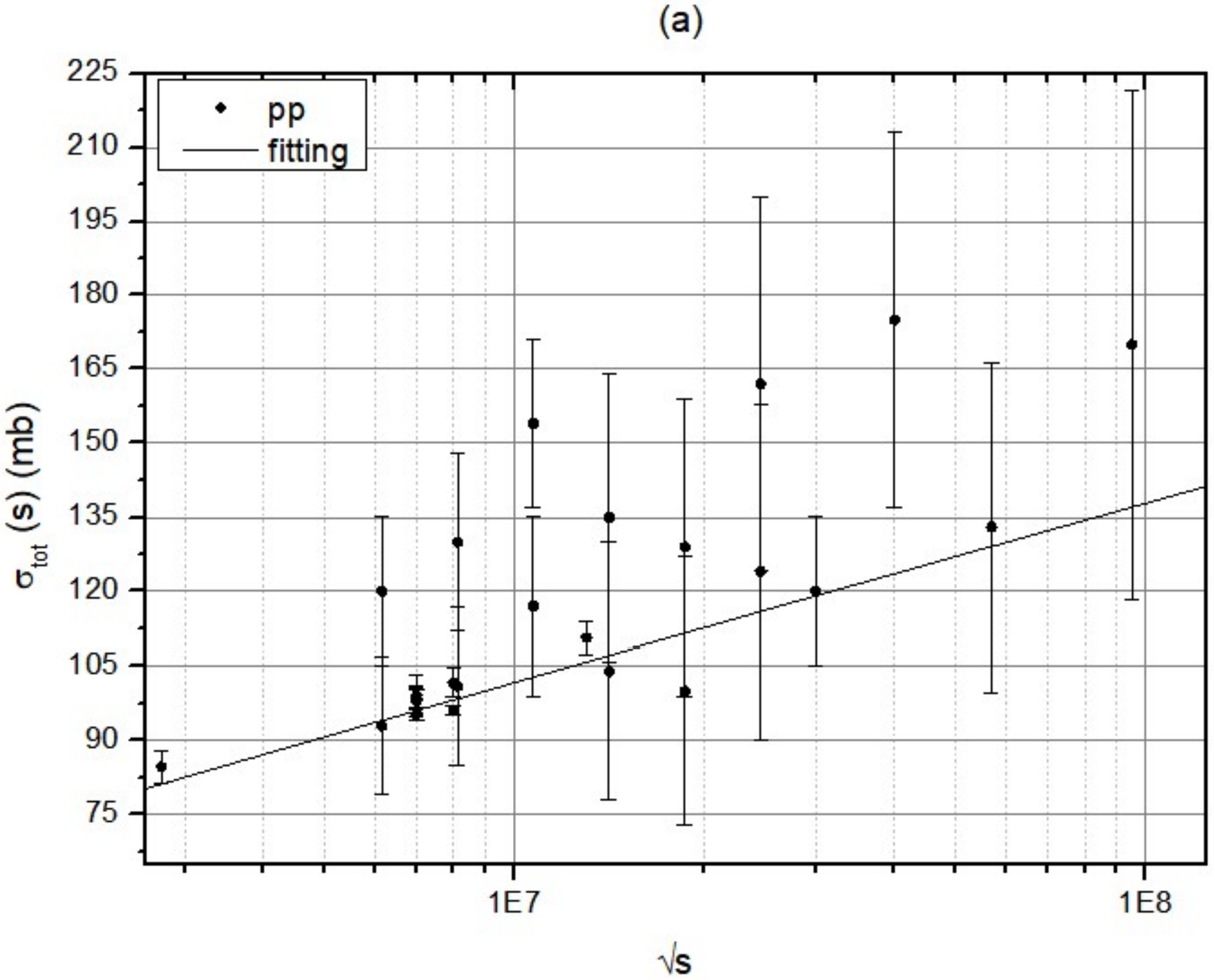}\vspace{-0.3cm}
\includegraphics[scale=0.3]{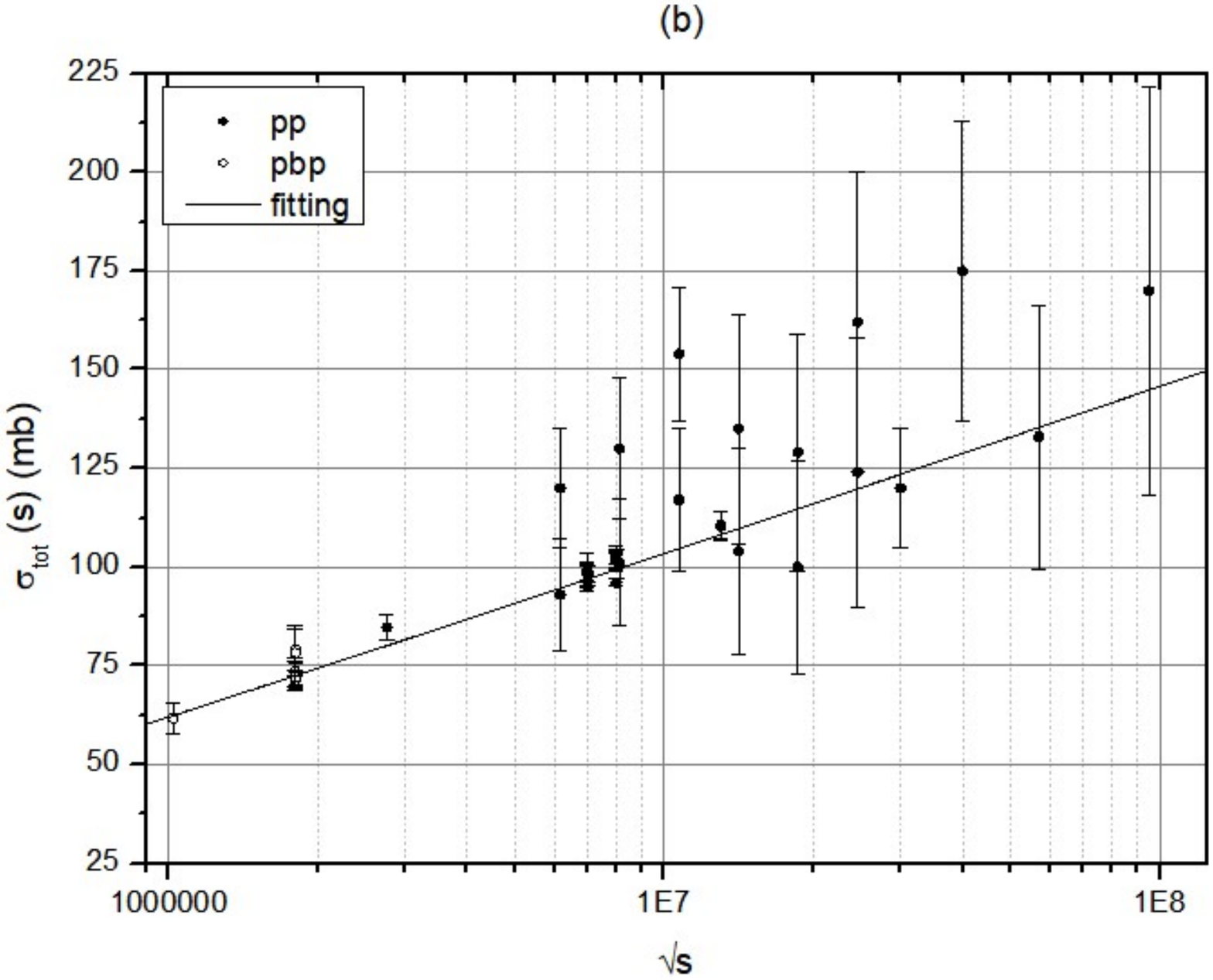} 
\vspace{-0.5cm}
\caption{The panel (a) shows the set 1 and the panel (b) the set 2. In both panels, the curve represents the fitting procedure using the parameterization (\ref{eq:sig_tot_1}). Experimental data are from \cite{PDG-PhysRev-D98-030001-2018} and $\sigma_{tot}^{pp}$ at $\sqrt{s}=2.76$ TeV is from \cite{G_Antchev_TOTEM_ Coll_Eur_Phys_J_C79_103_2019}.}
\label{fig:fig_1}}
\end{figure}

Figures \ref{fig:fig_1}(a) and \ref{fig:fig_1}(b) shows the fitting results for the set 1 and set 2, respectively. Table \ref{tab:table_2} displays the values of the fitting parameters. One can analyze these results considering the Froissart-Martin bound and the soft pomeron points of view. From the Froissart-Martin point of view, these values for the intercept ensure its non-saturation. The Froissart-Martin exponents obtained are 0.93$\pm$0.05 (set 1) and 1.05$\pm$0.05 (set 2), very below its expected saturation value. These intercepts represent a soft pomeron, indicating a process dominated by the exchange of a double pole - two gluon exchange. Furthermore, these intercepts are far from the Froissart-Martin bound saturation and below the usual hard pomeron prediction $\alpha_{\mathbb{P}}=1.4$ for diffractive processes \cite{P.V.Landshoff.hep.ph.0108156}.

Although simple, the parameterization (\ref{eq:sig_tot_1}) furnishes a reasonable statistical description of the data, as can be observed from the $\chi^2/ndf$ results shown in Table \ref{tab:table_2}. 

\begin{table}[h!]
\caption{\label{tab:table_2}Parameters obtained by using (\ref{eq:sig_tot_1}) in the fitting procedures.}
%\begin{indented}
%\item[]
\begin{tabular}{c | c | c | c}
\hline
set  ~&~ $\alpha_{\mathbb{P}}$ ~&~ $\beta$ (mb) ~&~$\chi^2/ndf$ \\ 
 \hline
1    ~&~ 0.93$\pm$0.19 ~&~ 10.04$\pm$4.58 ~&~ 1.48   \\
 \hline
2    ~&~ 1.05$\pm$0.05 ~&~ 7.54$\pm$0.92 ~&~ 1.26   \\
\hline
\end{tabular}
%\end{indented}
\end{table}

From these slopes, one can determine the mass of the particle by assuming a specific spin. Using the slope $\alpha'=0.25$ GeV$^{-2}$ from \cite{G.A.Jaroskiewicz.P.V.Landshoff.Phys.Rev.D.10.170.1974} for soft pomerons, one writes the linear Regge trajectory as $(|t|=m^2)$
\begin{eqnarray}\label{eq:regge_fit_1}
\alpha(m^2)=0.93+0.25m^2,
\end{eqnarray}

\noindent and
\begin{eqnarray}\label{eq:regge_fit_2}
\alpha(m^2)=1.05+0.25m^2,
\end{eqnarray}

\noindent and, for example, a particle with spin $J=\alpha(m^2)=2$ result in a particle with $m=2.07$ GeV and $m=1.95$ GeV, respectively. These results point out the existence of a soft pomeron with mass $\sim 2.0$ GeV.

 %On the other hand, using  $\alpha'=0.01$ GeV$^{-2}$ for hard pomerons \cite{P.V.Landshoff.hep.ph.0108156}, one obtains from these slopes \new{$m=10.34$ GeV and $m=9.75$ GeV}, respectively.

The relation between the internal entropy of the colliding hadron and the pomeron intercept may help to understand how the pomeron exchange contributes to the rise of the entropy. Recently, a simple scheme to calculate the hadron entropy has been proposed, assuming the Tsallis entropy as proportional to the real part of the profile function \cite{S.D.Campos.C.V.Moraes.V.A.Okorokov.Phys.Scrip.2019}.

%%%%%%%%%%%%%%%%%%%%%%%%%%%%%%%%%%%%%%%%%%%%%%%%%%%%%%%%%%%%%%%%%%%%%%%%
\section{Tsallis Entropy and the Regge Pole}\label{sec:ter}

The increasing correlation between the hadron internal constituents prevents the use of the Boltzmann entropy.  On the other hand, this strong correlation allows the use of the Tsallis entropy \cite{S.D.Campos.C.V.Moraes.V.A.Okorokov.Phys.Scrip.2019}. This entropy takes into account the strong interaction among quarks and gluons inside the hadron, using the information on the possible phase transition occurring in the total cross section experimental data at $\sqrt{s_c}$. The entropic index $w$ is replaced by the ratio $s/s_c$, allowing its physical interpretation in terms of the collision energy. The Tsallis entropy can be written in the impact parameter representation as \cite{S.D.Campos.C.V.Moraes.V.A.Okorokov.Phys.Scrip.2019}
\begin{eqnarray}\label{eq:tsallis_1}
 S_{T}(s,b)\approx \frac{1}{s/s_c-1}\bigl[1-\mathrm{Re}\Gamma(s,b)\bigr]^{2},
\end{eqnarray} 

\noindent where $\mathrm{Re}\Gamma(s,b)$ is the real part of the profile function, which can be connected with the Regge asymptotic amplitude (\ref{eq:asymp_2}) in the impact parameter space through a Fourier-Bessel transform. It is important to stress that in $b$-space the squared energy $s$ is fixed for each experiment.

The ratio $s/s_c$ can be written in a more general formulation as $(s/s_c)^{\alpha_1}$, where $\alpha_1$ can depend on the particle species and the collision type, for example. Currently, the information about the entropic index $w(s)=(s/s_c)^{\alpha_1}$ is limited to negative charged pions in proton-proton collisions, stating $\alpha_1=0.007$ \cite{EPJC-74-2785-2014}. On the other hand, there are studies in particle production processes indicating the rise of $w(s)$ with the collision energy \cite{EPJC-74-2785-2014,PLB-723-351-2013,AHEP-2016-9632126-2016,EPJA-53-102-2017}. Therefore, at present-day energies, $\alpha_1$ shows a weak $s$-dependence.

The Regge theory is based on the asymptotic behavior of the Legendre polynomial for $s\rightarrow\infty$, which means $s_c<\!\!<s$. In this situation, it is possible to find $\alpha_2$ satisfying the following approximation
\begin{eqnarray}\label{eq:factor}
\frac{1}{(s/s_c)^{\alpha_1} -1}\approx \frac{1}{\alpha_1\ln (s/s_c)^{\alpha_2}},
\end{eqnarray}

\noindent for $s>\!\!>s_c$. Therefore, the Tsallis entropy is written as
\begin{eqnarray}\label{eq:tsallis_2}
 S_{T}(s,b)\approx \frac{1}{\alpha_1\ln (s/s_c)^{\alpha_2}}[1-\mathrm{Re}\Gamma(s,b)\bigr]^{2}.
\end{eqnarray} 

\begin{figure}
\centering{
\includegraphics[scale=0.42]{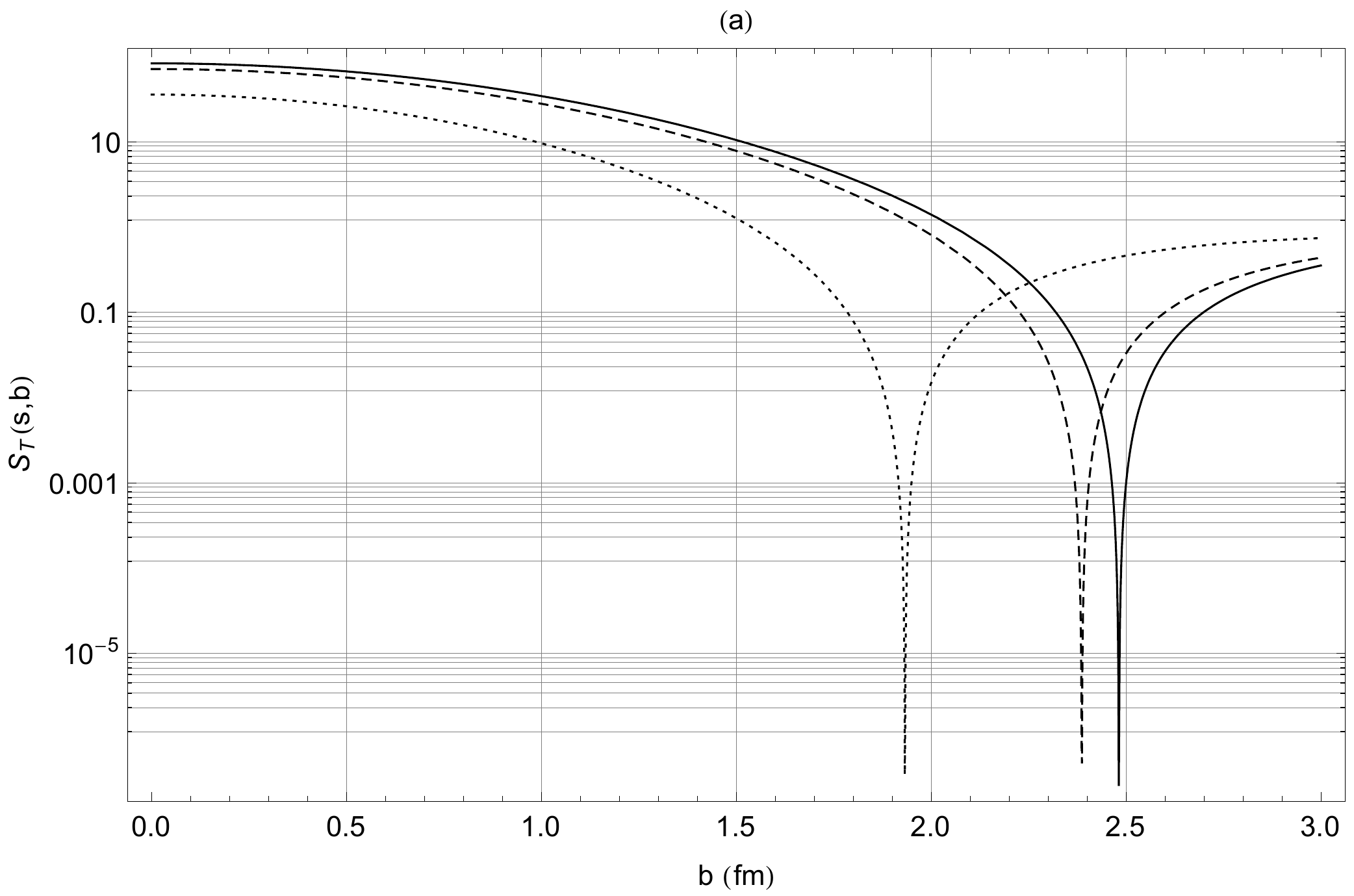}
\includegraphics[scale=0.42]{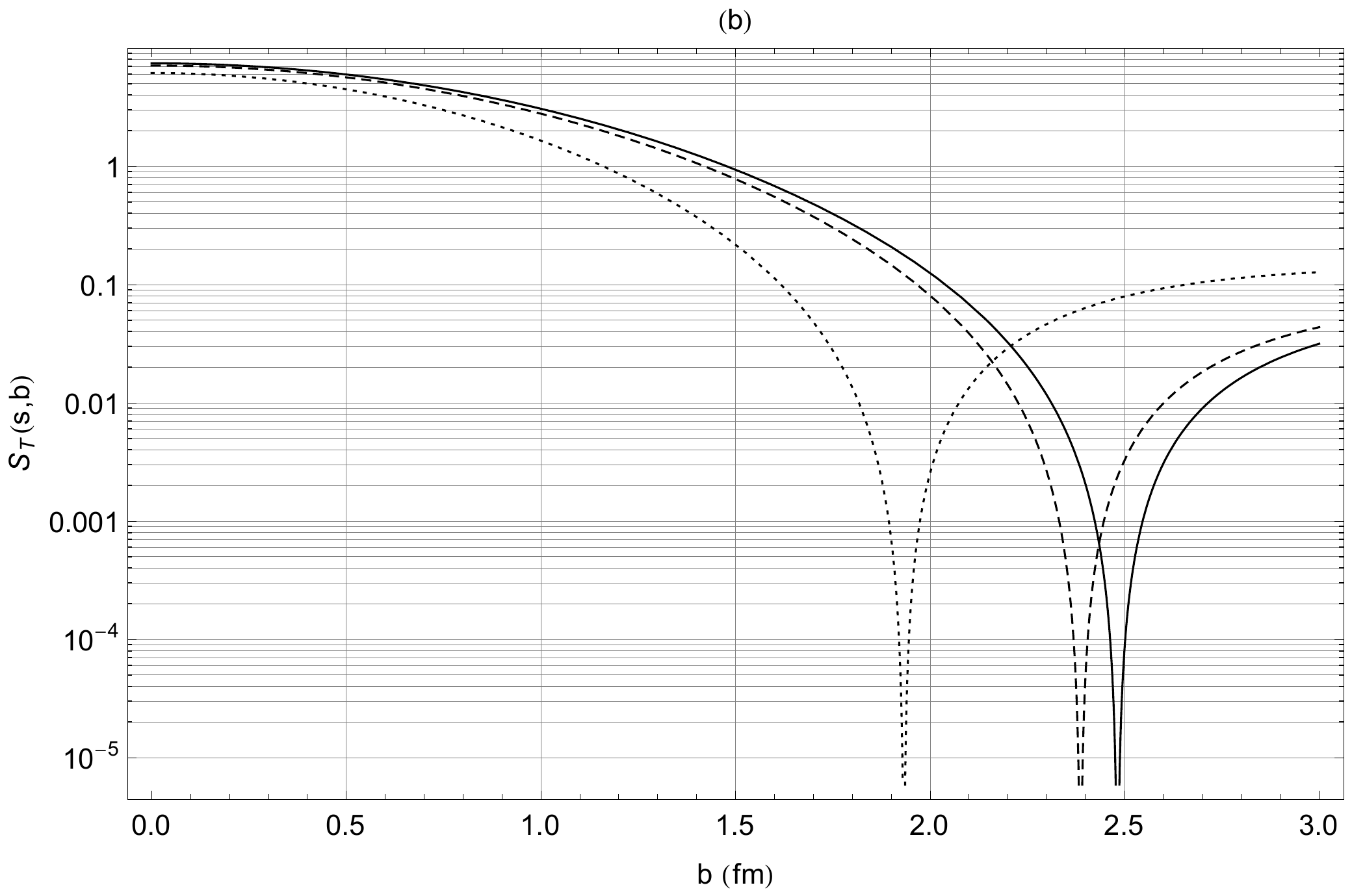} 
%\includegraphics[scale=0.4]{fig_2c.pdf}
%\includegraphics[scale=0.39]{fig_2d.pdf} 
%\vspace{-1.0cm}
\caption{The Tsallis entropy for the leading Regge pole using the critical energy $\sqrt{s_c}=25$ GeV. The solid-line is for $\sqrt{s}=14$ TeV, dashed-line is for $\sqrt{s}=7$ TeV and dotted-line is for $\sqrt{s}=600$ GeV. In (a) $\alpha_2=0.25$ and in (b) $\alpha_2=1.05$.}
\label{fig:fig_2}}
\end{figure}

The above result implies the entropy is positive for $s_c<\!\!<s$, as obtained in \cite{S.D.Campos.C.V.Moraes.V.A.Okorokov.Phys.Scrip.2019}. From the above discussion, one adopts without loss of generality $\alpha_1=1$,  since for $\alpha_1>0$ there is no changes in the entropy behavior \cite{S.D.Campos.C.V.Moraes.V.A.Okorokov.Phys.Scrip.2019}. Furthermore, the approximation (\ref{eq:tsallis_2}) may connect the Regge poles in the $t$-space with the behavior of the Tsallis entropy in $b$-space, as seen later. The $\mathrm{Re}\Gamma(s,b)$ can be obtained assuming the asymptotic amplitude (\ref{eq:asymp_2}), resulting in the Tsallis entropy generated by the leading Regge pole
\begin{eqnarray}\label{eq:profile_1}
 S_{T}(s,b)\approx \frac{1}{\ln (s/s_c)^{\alpha_2}} \left[1-\frac{e^{-\frac{b^2}{4\alpha'\ln(\ln s/s_c)}}\ln (s/s_c)^{\alpha_\mathbb{P}}}{2\alpha'\ln(\ln s/s_c)}\right]^2.
\end{eqnarray}

This result, despite constraints and approximations, is positive for $s_c<s$ and, except for the presence of $\alpha_2$, can be determined only by the leading Regge pole behavior. The role of $\alpha_2$ is analyzed in two situations. Firstly, assuming a $b$-independence, and later, a linear $b$-dependence.

The $b$-independence of $\alpha_2$ is displayed in the Figure \ref{fig:fig_2}, which show the Tsallis entropy for the leading Regge pole using (\ref{eq:profile_1}). Figure \ref{fig:fig_2}a is obtained by taking $\alpha_2=0.1$. Figure \ref{fig:fig_2}b shows the Tsallis entropy for $\alpha_2=\alpha_{\mathbb{P}}=1.05$. It is interesting to note that, for every $\alpha_2$, it is produced a non-physical entropy state outside the hadron. This result indicates the need for a cutoff for the Tsallis entropy in the Rege pole representation at the hadron edge, where one expects $S_T=0$. Thus, one considers the effective size of the hadron at $b_0$ where $S_T(s,b_0)=0$. Note that as the energy rises as well rises the largest value of the entropy.

Now, one adopts a linear $b$-dependence for $\alpha_2$. For the sake of simplicity, one uses for $\alpha_2$ a linear $b$-dependence applying the mnemonic rule $|t|\rightarrow b^{-2}$ into (\ref{eq:alpha_1}). Then is possible to connect this approach to the experimental results in $t$-space. Thus, one writes the following ansatz for $\alpha_2$
\begin{eqnarray}\label{eq:ansatz_b}
\alpha_2=\alpha(b)=\alpha_{\mathbb{P}}+\frac{\alpha'}{b^2}.
\end{eqnarray}

\begin{figure}[t]
\centering{
\includegraphics[scale=0.42]{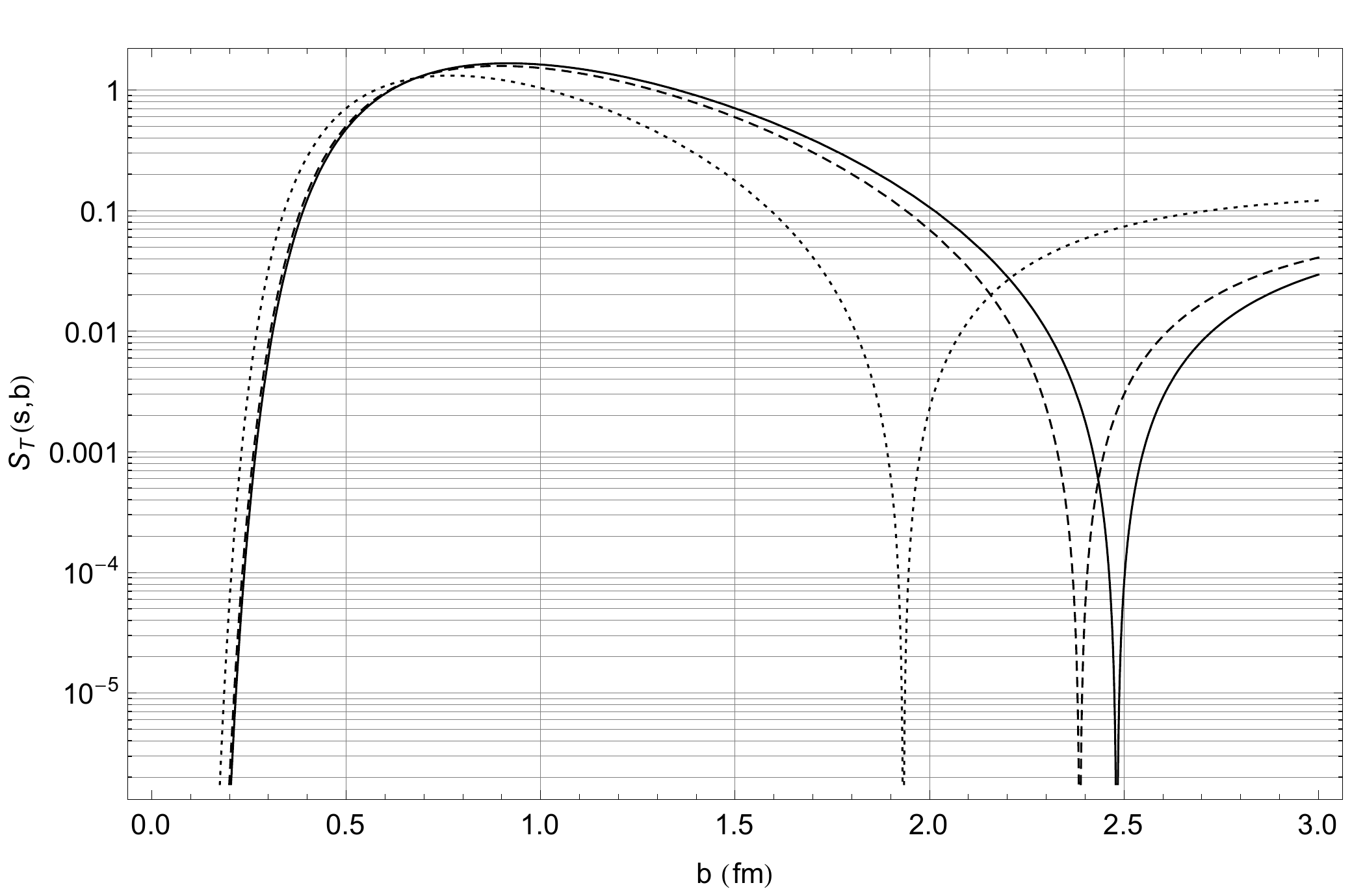}
%\includegraphics[scale=0.39]{fig_3b.pdf} 
%\includegraphics[scale=0.4]{fig_3c.pdf}
%\includegraphics[scale=0.39]{fig_3d.pdf} 
%\vspace{-1.0cm}
\caption{The Tsallis entropy for the leading Regge pole using the critical energy $\sqrt{s_c}=25.0$ GeV, $\alpha_\mathbb{P}=1.05$ and $\alpha'=0.25$ GeV$^{2}$. The solid-line is for $\sqrt{s}=14$ TeV, dashed-line is for $\sqrt{s}=7$ TeV and dotted-line is for $\sqrt{s}=600$ GeV.}
\label{fig:fig_3}}
\end{figure}

Figure \ref{fig:fig_3} shows the result for the entropy, considering (\ref{eq:ansatz_b}) and using the intercept obtained here and the slope taken from the literature. There is also a cutoff for the Tsallis entropy in this case. The main difference observed between these two Tsallis entropies is based on the adoption of $b$-dependence or $b$-independence for $\alpha_2$. First of all, assuming $\alpha_2$ as $b$-independent, then the entropy grows mainly near the hadron core, achieving its maximum at $b= 0$ fm. In this scenario, the main contribution for the entropy is given by the hadron internal constituents, located near the hadron core.

On the other hand, considering the $b$-dependent situation, one has $S_T\approx 0$ for $b\rightarrow 0$, as shown in the Figure \ref{fig:fig_3}. The main contribution for the entropy comes from the constituents located in the region $b\approx 0.5$ fm up to the hadron edge.  

These two results, based on the $\alpha_2$ behavior, introduces a novel perspective in the existence or not of the so-called gray area \cite{dremin_1,dremin_2}, also known as hollowness effect \cite{broniowski_arriola,alkin_martynov,anisovich_nikonov,troshin_tyurin_1,anisovich,troshin_tyurin_2,albacete_sotoontoso,arriola_broniowski}. As well-known, the absence of a maximum for the inelastic overlap function at $b=0$, for example, is the main consequence of such effect. In the picture introduced here, the existence of the hollowness effect is a consequence of the behavior of $\alpha_2$. It is important to stress that hollowness effect may exist for high $|t|$ values ($b\rightarrow 0$ fm), and taking into account the constraint $|t|<\!\!<s$. Therefore, this effect, if it exists, should occur only for $s$ above the TeV scale being negligible for $|t|\approx s$. In this situation, the approximation (\ref{eq:cos_2}) still holds, ensuring the validity of the above results. 

If $\alpha_2$ is independent of $b$, then there is no gray area since the entropy achieves its maximum at $b=0$ fm. However, if $\alpha_2=\alpha_2(b)$, then the gray area can emerge, being noted by the absence of the Tsallis entropy at $b=0$ fm, indicating a hollow in this region.

The Tsallis entropy obtained from the presented approach to the leading Regge pole, despite its simplicity, can help to distinguish if the main contribution for the collision is central or peripheral, based on the behavior of $\alpha_2$. A small value for the slope $\alpha'$ may indicate a small $t$-dependence of $\alpha_2$ (in this linear approach), which implies a contribution near the hadron core as being dominant over the peripheral one. %For example, assuming \new{$\alpha'=0.25$} GeV$^{-2}$, as performed above, then the hard pomeron picture is favored, indicating its presence may imply the non-existence of the hollowness effect.
A non-linear contribution for $\alpha_2$ can be considered (not presented here), resulting in a more complicated scenario. 

%%%%%%%%%%%%%%%%%%%%%%%%%%%%%%%%%%%%%%%%%%%%%%%%%%%%%%%%%%%%%%%%%%%%%%%%
\section{Critical Remarks}\label{sec:critical}%sect3

One obtains an approximation for the scattering angle, whose main result is a hadron-hadron total cross section arising from the asymptotic leading Regge pole, obeying the Froissart-Martin bound. It is important to stress, however, that $\sigma_{tot}$ used in the present analysis comes from a scattering amplitude without subtractions, which implies that Cauchy integral theorem may not be satisfied. Nonetheless, this is a natural consequence of the Regge approach for wave expansion, which result in a scattering amplitude not necessarily satisfying such a theorem. The advantage of the present approach over the well-known Donnachie and Landshoff model, for example, is the preservation of unitarity, as well as the Froissart-Martin bound.

The introduction of subtractions, however, can be easily performed preserving the Cauchy integral theorem, unitarity, and Froissart-Martin bound, resulting in $\sigma_{tot}(s)\approx s^{\gamma}\ln (s/s_c)^{\alpha_\mathbb{P}(0)}$, $\gamma<0$. On the other hand, the natural consequence of that is the decreasing of the total cross section above some $\sqrt{s_d}$, as $s\rightarrow\infty$. Of course, the pomeron picture obtained cannot be identified with the simple Regge trajectory, being more similar to a Regge cut term. The analysis of this case will be performed elsewhere.

The approximation implemented is valid for $0\leq \cos \theta \leq 1$ and $|t|<\!\!< s$, which includes the forward collision case as well as the cases where the inequality holds. In a first moment, the $pp$ total cross section was analyzed only for $\sqrt{s}\geq 1.0$ TeV, including the cosmic-ray data. The fitting procedure results in a pomeron intercept $\alpha_{\mathbb{P}}=0.93\pm 0.19$, indicating the presence of a soft pomeron. In a second moment, the experimental data for $\sigma_{tot}^{p\bar{p}}$ above 1.0 TeV were added to the $pp$ data in an attempt to compensate the absence of $pp$ experimental data in the energy range $1.0$ TeV$<\sqrt{s}<2.0$ TeV. The main goal of this procedure was to enhance the statistical description of the data resulting in $\alpha_{\mathbb{P}}=1.05\pm 0.05$. These values for $\alpha_{\mathbb{P}}$ indicates the dominance of two gluon exchange and a very slow rising $\sigma_{tot}^{pp}$ above $1.0$ TeV.

%The high $\chi^2/ndf$ values, however, indicates the toy-model is not sufficient to describe the experimental data in the proposed range. In physical terms, one can say the toy-model introduced is not taking into account the non-leading (secondary) pole contribution.

Using the proposed parameterization for the fitting procedure, one obtains the Tsallis entropy in the $b$-space. A simple mathematical approximation was implemented, allowing the use of a logarithmic representation for the entropic index $w$. Due to this approximation, emerges a free parameter analyzed as being $b$-independent and $b$-dependent. In the first case, the entropy reveals that the main contribution to the elastic scattering pattern is central since the entropy achieves its maximum at $b=0$ fm. In the later case, one assumes $\alpha(b)$ given by the simple change of $|t|\rightarrow b^{-2}$ in the linear form of $\alpha(t)$. This assumption results in a Tsallis entropy mainly generated in the region $b\neq 0$ up to the hadron edge. Of course, the mnemonic relation $b^2\rightarrow |t|^{-1}$ implies $b=0$ only for $|t|\rightarrow\infty$, which cannot occurs without violation of the mandatory condition $|t|<\!\!<s$. However, if one considers the present-day range for $|t|$, then the obtained results are valid as an approximation, working better for $\sqrt{s}$ above the TeV scale. Indeed, the momentum transfer for the $pp$ collisions is restricted to $|t|<15$ GeV$^2<\!\!<s$ resulting in $b\approx 0.05$ fm, very close to the forward collision. 

It is interesting to note that the leading Regge pole is obtained in asymptotic conditions ($s\rightarrow\infty$). Thus, it would be expected to work only for very high energies. Yet, it is surprising that it works in the energy range one has nowadays, far from the concept of asymptotia. The Pomeranchuk theorem, in the original and modified versions, had predicted the existence of an exchange particle that does not differ particle-particle and particle-antiparticle scattering at very high energies. This theorem also seems to be valid at the present-day energy scale. At a low energy scale, on the other hand, the total cross section for $pp$ and $p\bar{p}$ have different patterns. Furthermore, the presence of the odderon seems inevitable to fill up the dip in the $pp$ and $p\bar{p}$ differential cross section \cite{L.L.Jenkovszky.A.I.Lengyel.D.I.Lontkovskyi.Int.J.Mod.Phys.A26.4755.2011}. %In this sense, the present work seems to reinforce that result, at least for the \new{$\sqrt{s}>1.0$} TeV range, since this region may have contributions coming from the odderon exchange. However, it is important to stress that the energy range below $\sqrt{s_c}$ was not taken into account in the fitting procedures performed here. 

The BFKL equation is the perturbative mechanism driving the growth of the total cross section \cite{E.A. Kuraev.L.N.Lipatov.V.S.Fadin.Sov.Phys.JETP.45.199.1977,I.I.Balitsky.L.N.Lipatov.Sov.J.Nucl.Phys.28.822.1978}, and the  BFKL pomeron can be calculated from its series expansion \cite{R.Kirschner.L.N.Lipatov.Z.Phys.C45.477.1990}.
%\begin{eqnarray}\label{bfkl_1}
%\alpha_\mathbb{P}(\mu^2)=1+\frac{12\ln2}{\pi}\alpha_s(\mu)\gamma(\mu),
%\end{eqnarray}
%\noindent where
%\begin{eqnarray}\label{bfkl_2}
%\gamma(\mu)= %1-\alpha_s(\mu)^{2/3}\left(\frac{7\zeta(3)}{2\ln2}\right)^{1/3}\left( %\frac{3(3/4+n_r}{33-2n_f} \right)^{2/3}+...  
%\end{eqnarray}
%The QCD running coupling constant is written as
%\begin{eqnarray}\label{running_1}
%\alpha_s(\mu^2)=\frac{1}{\beta_0\ln(\mu^2/\Lambda_{QCD}^2)},
%\end{eqnarray}
%\noindent and $\beta_0=(33-2n_f)/12\pi$ is the 1-loop function. 
The hard pomeron describes processes dominated by small transverse distance, whereas the soft pomeron describes large transverse distances ($\sim$ the proton radius). Although the BFKL equation was not created for the soft pomeron, there are interpolating methods being developed about a possible BFKL approach for the soft pomeron \cite{J.Bartels.C.Contreras.G.P.Vacca.JHEP.01.004.2019} (and references therein). The comparison between the results obtained here and this BFKL approach will be presented elsewhere. 
%As a prediction, using} the masses obtained above for the $\alpha'=0.25$ GeV$^{-1}$, one gets both the running coupling constant and the BFKL pomeron intercept. Therefore, assuming $n_r=0$, $n_f=6$, $\Lambda_{QCD}^2=0.332$ GeV$^2$ the resulting $\alpha_s(|t|)$ are: \new{$\alpha_s=0.123$ and $m=1.29$ GeV for $\alpha_{\mathbb{P}}=0.93$, and \new{$\alpha_s=0.127$ and $m=1.30$ GeV for $\alpha_{\mathbb{P}}=1.05$}. These values are below those calculated by the linear approach (\ref{eq:alpha_1}), which may state that BFKL is not sufficient to explain the hard pomeron. Nonetheless, the contrary situation may be the correct one: the linear approach is not able to describe the hard pomeron trajectory.}

%The occurrence of the hollowness effect cannot be ensured by the present fitting toy-model. Nonetheless, it can be analyzed by the approach to the leading Regge pole presented here and, therefore, its physical origin may depend on the hard pomeron presence as the exchange particle in the hadron-hadron collision. 

It is important to stress that the term soft pomeron emerges in the context of $\sigma_{tot}(s)\sim s^{\alpha(0)}$, for $\alpha_{\mathbb{P}}\gtrapprox 1$. The hard pomeron appears when the Regge phenomenology is used, by analogy, to explain the small-$x$ data for $F_2(x,Q^2)$. Thus, applying the present formalism to diffractive processes may result in an intercept higher than the one obtained here.

As a complimentary question, one must to say that the pomeron intercept depends on the renormalization scheme and scale for the running coupling constant \cite{E.A.Kuraev.L.N.Lipatov.V.S.Fadin.Sov.Phys.JETP.44.443.1976,L.N.Lipatov.Sov.J.Nucl.Phys.23.338.1976,Ya.Ya.Balitzkij.L.N.Lipatov.Sov.J.Nucl.Phys.28.822.1978}. However, the Regge pole should not depend on this arbitrary choice, since it has a phenomenological basis, the Chew-Frautschi plot. Thus, the definition of a renormalization group for the Regge theory is an important question to be solved in the future.

%Performing a suitable change in the scattering angle near the forward direction, it is possible to express total cross section due to the leading Regge pole as obeying the Froissart-Martin bound. The total cross section for $pp$ elastic scattering is, then, used to obtain the pomeron intercept $\alpha_{\mathcal{P}}=1.77\pm0.02$, implying in a hard pomeron picture. On the other hand, the Tsallis entropy can also be connected with the leading Regge pole, showing that this kind of entropy is generated by the leading Regge pole.

%Using the Helmholtz free energy and a suitable change of variables, one can obtain the Callan-Symanzik equation. Regge pole picture can also be used...

%%%%%%%%%%%%%%%%%%%%%%%%%%%%%%%%%%%%%%%%%%%%%%%%%%%%%%%%%%%%%%%%%%%%%%%%
\section*{Acknowledgments}

SDC thanks to UFSCar by the financial support.

\end{document}